# A MOBILE WEB FOR ENHANCING STATISTICS AND MATHEMATICS EDUCATION

Jamie Lentin,[1] Anna H. Jonsdottir,[2] David Stern,[3] Victoria Mokua,[3] Eva Dogg Steingrimsdottir,[2]
Magnea Run Vignisdottir[2] and Gunnar Stefansson[2]
[1]Shuttle Thread, Manchester, England
[2]University of Iceland; Science Institute, Taeknigardur Dunhanga 5; 107 Reykjavik; Iceland
[3]Maseno University, Private Bag, Maseno, Kenya
Contact email: gstefans@gmail.com

**Abstract:** *A freely available educational application (a mobile website) is presented. This provides access to educational material and drilling on selected topics within mathematics and statistics with an emphasis on tablets and mobile phones. The application adapts to the student's performance, selecting from easy to difficult questions, or older material etc. These adaptations are based on statistical models and analyses of data from testing precursors of the system within several courses, from calculus and introductory statistics through multiple linear regression. The application can be used in both on-line and off-line modes. The behavior of the application is determined by parameters, the effects of which can be estimated statistically. Results presented include analyses of how the internal algorithms relate to passing a course and general incremental improvement in knowledge during a semester.*

## INTRODUCTION

Many on-line drilling systems exist, with some specially designed for a specific topic whereas others are general in nature. The "tutor-web" is a general system for drilling students in addition to storing general educational content. This system has been tested and used by over 2000 students, mostly in introductory courses on statistics and mathematics. Design principles include content modularity, open source software, creative commons texts and drilling exercises which are freely accessible to all students without regard to their physical location or whether they are registered into any school or university. Earlier versions of the system have been written for various platforms and in different programming languages (Stefansson, 2004; Jonsdottir, Jakobsdottir & Stefansson, 2014). These have been used to test various concepts and have forged the basis for the algorithms implemented here.

The system is made freely available and primarily intended for learning, not mere evaluation. Therefore students are encouraged to continue using the drilling system until they have achieved expertise on the topic in question. For this reason there is no limit on how long students can request new questions within the drilling system. Research efforts have therefore concentrated on ensuring that the system actually entices students to continue until a high level of expertise is obtained. Since research shows students continuing until a high grade is achieved, the grading scheme needs to be formulated so that a high grade reflects a high level of expertise.

At the core of the tutor-web is the use of formative assessment in drills. Formative assessment has been found to be effective in building knowledge in students (Black & William, 1998). After a student answers a drill, a step-by-step solution is normally provided so the student can understand where they went wrong.

A drilling system is, by nature, different from a computer-aided-testing (CAT) system. It has been demonstrated that learning occurs during the typical tutor-web practice session (Stefansson & Sigurdardottir, 2011), whereas the Item Response Theory (IRT) used in CAT uses models, which do not permit learning. For this reason a drilling system will allocate items (questions), which are typically easy initially but become more difficult as the student's grade increases.

It has been seen that students using on-line drills try to work very hard towards a high grade when given the option to do so (Stefansson & Sigurdardottir, 2011), but potentially with extensive guessing. In combination with a grading scheme based on the last 8 answers, this results in drill grades, which can be far too high when compared to performance on an exam (Stefansson, 2004). It has therefore been proposed that one could implement a timeout option so that a high grade ensures not only that the student has the capability to eventually solve a problem but has the expertise to solve it quickly (Jonsdottir and Stefansson, 2013). One can also implement a lecture grade, which is a taper of recent grades with a tail, which becomes longer with extended guessing (as proposed in Desjardins et al., 2014). As shown below, it turns out that implementing the timeout and longer tail has a considerable effect.

THE MOBILE TUTOR-WEB DRILLING SYSTEM

Drill questions are organized by course, tutorial and lecture. A student, upon visiting tutor-web and logging in, can explore the courses available and find a lecture they wish to load onto their computer/mobile. Alternatively they can proceed directly to the drill interface, where they get a choice of already-loaded lectures to study. Either way, once a lecture is chosen they can start working through questions. The interface they see is shown in the screenshot in Figure 1.

A student is presented with a question, selected based upon their current grade, and a choice of answers, both of which can involve TeX equations as well an image. They then have to choose one of the answers within a specified time- there is a countdown timer near the bottom of the screen. The answers are displayed in a random order to avoid learning where correct answers are placed. Once an answer has been selected they will then be shown whether their answer is correct or not, and an explanation as to why this is the correct answer. They can also see their current grade, and how many questions for this lecture they have answered.

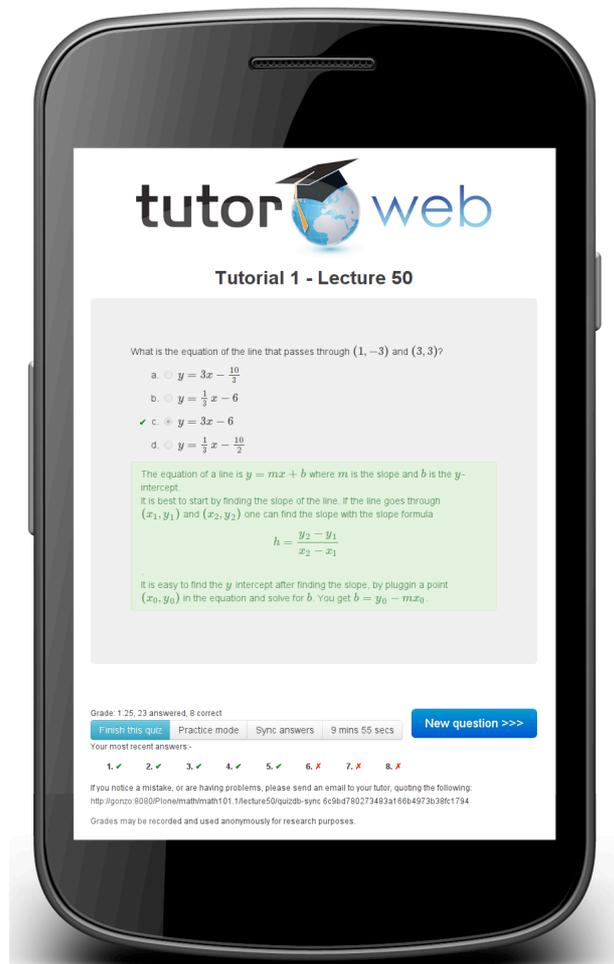

**Fig. 1.** Screen shot of mobile tutor-web question after student has submitted a correct answer. Note the detailed explanation, which has subsequently appeared below the question.

The drill interface runs entirely on the student's device using HTML and Javascript, which means it is capable of working on any modern mobile, tablet or desktop computer. It utilises AppCache and LocalStorage to store the code and the question data on the student's device, this means that the interface remains quick, and can even work without any Internet connection. Answers are simply stored locally until there is an internet connection to send them back to the tutor-web server.

The server is based on the Plone CMS and MySQL. Plone provides user/class management, as well as storing the banks of questions. Questions can be imported from TeX files as well as entered and edited manually. Once a student chooses a lecture, then the drill interface asks the server for up to 100 randomly allocated questions from the lecture. The random allocation both ensures that we do not fill the device with too many questions and gives an amount of security, as each student will not be answering the same questions. We also give questions long sparse references that are tied to individual students, so a student cannot download an entire question bank by guessing IDs, or download a question allocated for another student.

Periodically, the drill interface will send any answers to questions back to the tutor-web server. A class tutor will get to see the progress of their class from an administration interface, and the raw answers will be available via MySQL for further analysis.

The algorithms used to set the timeout are the first implementation of those proposed in Jonsdottir and Stefansson (2013) and described in more detail in Desjardins et al. (2014). Basically, the timeout is in the shape of an inverse dome curve, with a minimum time set to correspond to some grade within the lecture. In this manner, beginning or struggling students do not get affected by the timeout, and the timeout does not affect the most difficult or time-consuming questions. However, the students can not get a high grade or proceed to the most difficult questions without passing through the timeout set at the intermediate-level grades. These settings are determined by a parametric function flexible enough to also allow for no (or high, constant) timeout.

Earlier versions of the tutor-web (non-mobile) just used the most recent 8 answers to compute a grade within each lecture. It became clear from earlier experiments that (a) students could fairly easily guess their way to what they considered an adequate grade and (b) students tended to quit if they answered a questions incorrectly after 7 correct in a row. A taper was therefore implemented in the mobile web version and used for grades in the following analysis. This initial taper is simply an average of the most recent responses, starting with the most recent 8, but the tail gets longer (n/2) as the number of responses (n) increases above 16, but only up to a maximum of 30 questions in the grade (achieved at n=60 or more). In principle the weights given to the answers could be described by a parametric function, with the effects of different parameter settings to be estimated with formal experimentation. As described in Desjardins et al. (2014), the resulting tutor-web grade should become a better predictor of whether the students pass or fail when the grading scheme includes a longer tail. Desjardins et al. (2014) also include a simple ROC-type analysis to conclude that the combined effects of a timeout+taper appear to correspond to a more internally consistent grading mechanism, with a more elaborate analysis is given below. It is therefore suggested that one should test the effect of other down weighting schemes and possibly also the effect of a higher weight on the very last answer, to entice students to continue.

USES

Currently the tutor-web system is mainly a support tool, used to supplement education in the classroom. The most elaborate tests of the mobile system described here (i.e. at http://mobile.tutor-web.net) have been in an undergraduate setting, i.e. in one large calculus course (450 entrants) and one large introductory statistics course (250 entrants). In addition the mobile tutor-web has been used for secondary school (high school) mathematics. The system has been tested and used in several high schools (secondary schools) in Iceland, for large and small classes at the University of Iceland and for a very large class at Maseno University in Kenya. Additional uses have been more sporadic, but the system is freely accessible and no formal record is maintained of the user's whereabouts except when an instructor decides to use it with a class.

Course content for a graduate remedial calculus/programming course for statistical modellers also exists. This overview course demonstrates a particularly useful aspect of an open system such as the tutor-web. It was initially developed for graduate students at the university of Iceland, in fields other than mathematics and engineering. Such students frequently lack the basis for taking advanced statistics courses and hence this course was developed as that base. The content therefore covers introductory calculus, linear algebra and multivariate calculus along with computer programming. Although it covers the basic theory, it has a very applied slant and all examples are from statistics. The course has subsequently been used as an entry criterion for applicants to a PhD programme: Applicants have been asked to take this course remotely, in addition to satisfying other formal criteria for entering a PhD study. This demonstrates how on-line content, with drills, can be used as a supplement to in-class teaching, for remote learning and as remedial material for students lacking prerequisite background knowledge.

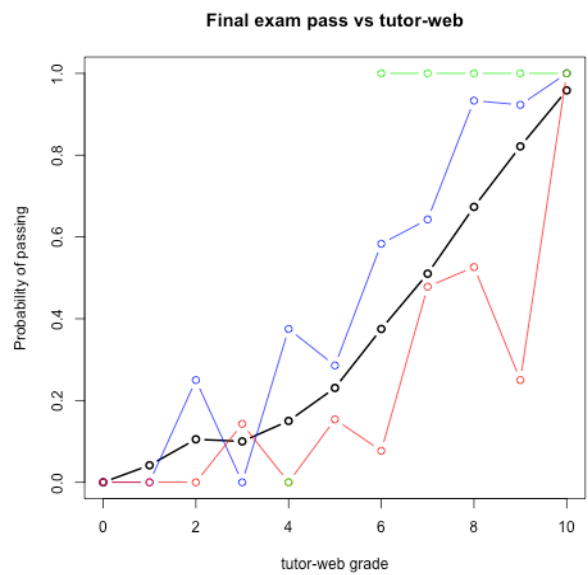

**Fig. 2.** Results from experiments in a 2013 introductory calculus class at the University of Iceland. The black curve describes the probability of passing the final exam, as a function of the overall grade from tutor-web work during the semester. Also shown are the same results, split up according to performance on a status exam, given at the beginning of the semester (red=poor, blue=medium performance and green=excellent).

Earlier versions of the tutor-web (at http://tutor-web.net) have more content (which will migrate to the mobile tutor-web). This includes courses on linear algebra, multiple linear regression, analysis of variance, survival analysis, earth science and marine population dynamics. Some of these courses are available in more than one language (English and Icelandic).

EXPERIMENTAL RESULTS

In addition to the results given below, results from various experiments using earlier (non-mobile) versions of the tutor-web system are reported in the cited papers. These range from demonstrations of student progress while using the system to analyses of stopping times, which show how properties of the grading scheme strongly affect student behavior while using the system. Several in-class surveys reflect a general student satisfaction with the methodology while also demonstrating that an on-line system alone is not enough: Students want a combination of the drilling system which gives immediate feedback as well as regular, marked, homework. Earlier analyses which show that students continue until they receive a high grade, but the high grade did not correspond well to expertise, have led to revisions of the grading scheme so the version implemented in the mobile web much better reflects knowledge (as measured by a final exam) than did earlier grading schemes.

A system like the tutor-web has considerable potential for low-income regions (see http://tutor-web.tumblr.com/post/59494811247/web-assisted-education-in-kenya for an overview of experiments in Kenya). In fact, student surveys in Kenya give comments similar to the ones received in Iceland, most highly positive. The two comments from Kenya, "Doing maths online was the best experience I ever had with maths" and it "was interesting to do maths online" contrasted starkly with a previous survey for the same service course which produced comments of almost universal discontent. There are substantial differences to the circumstances but the fact that "I wished to do more" is typical of the responses obtained in either region shows the universality of the benefits for motivation.

The importance of moving towards a mobile based system in countries like Kenya cannot be overstated. The number of students having access to smart phones is exploding and this opens up a whole new set of possibilities by freeing students from over-crowded and difficult to maintain computer labs. Results with similar solutions at school and diploma level (Manyala, Mbasu, Stern & Stern, 2014) have shown the potential to impact performance and motivation across a broad range of the academic spectrum. The move to mobile opens up the possibility for these systems to scale both within and beyond the classroom; these opportunities have only just started to be explored (see http://momaths.nokia.com/).

Fig. 2 is based on data from 200+ students and compares the performance on a final exam to the performance in the mobile tutor-web (black curve). A strong relationship is clear in the figure, where the probability of passing goes from negligible to almost 100% as the web-based performance is taken from 0 to a perfect grade of 10. This curve alone can not claim to describe a cause and effect relationship since good students entering the course will be at the high end of both graphs. A status exam is available, and this is given at the beginning of the course. The students' grades on this exam range from extremely poor through medium to quite high and when the data are split into these categories (red, blue and green, respectively), the same pattern is observed.

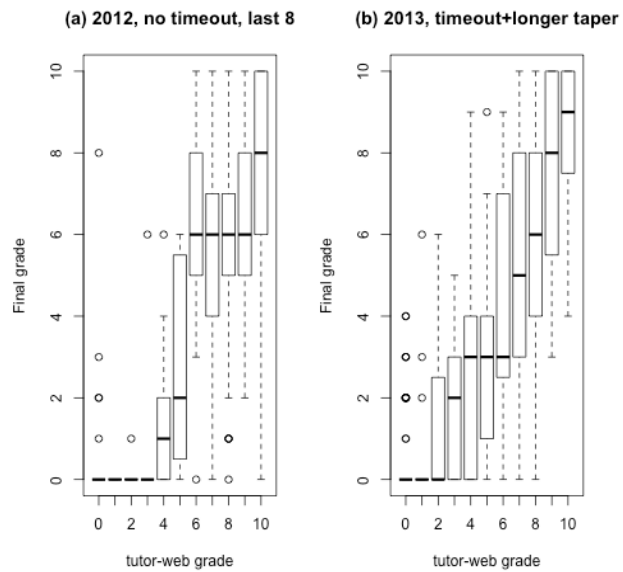

**Fig. 3.** Comparisons of results from experiments in calculus classes in 2012 and 2013 at the University of Iceland. In both panels, the y axis shows the grade on the final exam, as a function of the overall grade from tutor-web work during the semester.

It is of greater interest to compare directly the effects of the revisions involved in the mobile web implementation to earlier results. The mobile web implementation included the implementation of a timeout option as well as a change in the tapering used for computing a grade.

The effects of these changes are evaluated in Fig. 3 by comparing the relationships between the final grade and the tutor-web grade for the two systems. This is done using data from the introductory calculus course in 2012 (old system, fixed taper of length 8 and no timeout) to the data from 2013 (mobile web, variable taper of length up to 30 and timeout). The important conclusion from these data is that in 2012 there is no obvious relationship between tutor-web performance and the final grade, in the tutor-web grade range 6-9. This was an important failing of the internal grading system, since the old grade did not give the students any indication whether they might have a problem or not. The revised system seems to be a much better performance indicator (though clearly it would be very useful to reduce the variability in the prediction, as shown by the width of the boxes).

CONCLUSIONS

When compared with the previous versions of the tutor-web (Jonsdottir et al., 2014), the mobile tutor-web is much faster and more visually appealing in all regards. All earlier versions also implemented equations as static images whereas the MathJax implementation (http://www.mathjax.org/) enables zooming into and out of the equations. This is an important feature, both for tablets per se due to the screen size and also for complicated equations on any screen.

The mobile tutor-web includes much better user interfaces for students and their instructors than previous systems, and gives more flexibility to the content provider in terms of parameters controlling the behavior of the grading scheme and the item allocation. The mobile version does not yet, however, provide an easy link for the content provider to store and edit other educational material in raw form so this can only be provided in PDF format. Although this is no loss to the student user, an important motive for the tutor-web was cooperation among instructors to provide content and this needs to be re-implemented in the database underlying the mobile tutor-web.

ACKNOWLEDGEMENTS

The research leading to the results in this paper has received funding from the European Union Seventh Framework Programme (FP7/2007-2013) under grant agreement 613571 - MareFrame.